\PassOptionsToPackage{unicode}{hyperref}
\PassOptionsToPackage{hyphens}{url}
\PassOptionsToPackage{dvipsnames,svgnames,x11names}{xcolor}
\documentclass[
]{article}
\usepackage{amsmath,amssymb}
\usepackage{lmodern}
\usepackage{iftex}
\ifPDFTeX
  \usepackage[T1]{fontenc}
  \usepackage[utf8]{inputenc}
  \usepackage{textcomp} 
\else 
  \usepackage{unicode-math}
  \defaultfontfeatures{Scale=MatchLowercase}
  \defaultfontfeatures[\rmfamily]{Ligatures=TeX,Scale=1}
\fi
\IfFileExists{upquote.sty}{\usepackage{upquote}}{}
\IfFileExists{microtype.sty}{
  \usepackage[]{microtype}
  \UseMicrotypeSet[protrusion]{basicmath} 
}{}
\makeatletter
\@ifundefined{KOMAClassName}{
  \IfFileExists{parskip.sty}{%
    \usepackage{parskip}
  }{
    \setlength{\parindent}{0pt}
    \setlength{\parskip}{6pt plus 2pt minus 1pt}}
}{
  \KOMAoptions{parskip=half}}
\makeatother
\usepackage{xcolor}
\usepackage{graphicx}
\makeatletter
\def\maxwidth{\ifdim\Gin@nat@width>\linewidth\linewidth\else\Gin@nat@width\fi}
\def\maxheight{\ifdim\Gin@nat@height>\textheight\textheight\else\Gin@nat@height\fi}
\makeatother
\setkeys{Gin}{width=\maxwidth,height=\maxheight,keepaspectratio}
\makeatletter
\def\fps@figure{htbp}
\makeatother
\newlength{\cslhangindent}
\newlength{\csllabelwidth}
\newlength{\cslentryspacingunit} 
\newenvironment{CSLReferences}[2] 
 {
  \setlength{\parindent}{0pt}
  \ifodd #1
  \let\oldpar\par
  \def\par{\hangindent=\cslhangindent\oldpar}
  \fi
  \setlength{\parskip}{#2\cslentryspacingunit}
 }%
 {}
\usepackage{calc}

\ifLuaTeX
\usepackage[bidi=basic]{babel}
\else
\usepackage[bidi=default]{babel}
\fi
\babelprovide[main,import]{american}

\def\languageshorthands#1{}
\ifLuaTeX
  \usepackage{selnolig}  
\fi
\IfFileExists{bookmark.sty}{\usepackage{bookmark}}{\usepackage{hyperref}}
\IfFileExists{xurl.sty}{\usepackage{xurl}}{} 
\hypersetup{
  pdftitle={ShOpt.jl: A Julia Package for Empirical Point Spread
Function Characterization of JWST NIRCam Data},
  pdfauthor={Edward Berman, Jacqueline McCleary},
  pdflang={en-US},
  colorlinks=true,
  linkcolor={Maroon},
  filecolor={Maroon},
  citecolor={Blue},
  urlcolor={Blue},
  pdfcreator={LaTeX via pandoc}}

\title{ShOpt.jl: A Julia Package for Empirical Point Spread Function
Characterization of JWST NIRCam Data}

\usepackage[affil-it]{authblk}
\usepackage{orcidlink}
\author[1%
  \ensuremath\mathparagraph]{Edward Berman%
    \,\orcidlink{0000-0002-8494-3123}\,%
    }
\author[1%
  \ensuremath\mathparagraph]{Jacqueline McCleary%
    \,\orcidlink{0000-0002-9883-7460}\,%
    }

\affil[1]{Northeastern University, USA}
\affil[$\mathparagraph$]{Corresponding author}
\date{Submitted: 23 August 2023, Accepted: 25 January 2024}

\begin{document}
\maketitle

\hypertarget{summary}{%
\section{Summary}\label{summary}}

When astronomers capture images of the night sky, several factors --
ranging from diffraction and optical aberrations to atmospheric
turbulence and telescope jitter -- affect the incoming light. The
resulting distortion and blurring are summarized in the image's point
spread function (PSF), the response of an optical system to an idealized
point source. The PSF can obscure or even mimic the astronomical signal
of interest, making its accurate characterization essential. By
effectively modeling the PSF, we can predict image distortions at any
location and proceed to deconvolve the PSF, ultimately reconstructing
distortion-free images.

The PSF characterization methods used by astronomers fall into two main
classes: forward-modeling approaches, which use physical optics
propagation based on models of the optics, and empirical approaches,
which use stars as fixed points to model and interpolate the PSF across
the rest of the image. (Stars are essentially point sources before their
light passes through the atmosphere (when observing from the ground) and
telescope, so the shape and size of their surface brightness profiles
define the PSF at that location.) Empirical PSF characterization
proceeds by first cataloging the observed stars, separating the catalog
into validation and training samples, and interpolating the training
stars across the field of view of the camera. After training, the PSF
model can be validated by comparing the reserved stars to the PSF
model's prediction.

Shear Optimization with \texttt{ShOpt.jl} introduces modern techniques,
tailored to James Webb Space Telescope (JWST) NIRCam imaging, for
empirical PSF characterization across the field of view. ShOpt has two
modes of operation: approximating stars with analytic profiles, and a
more realistic pixel-level representation. Both modes take as input a
catalog with image cutouts -- or ``vignettes'' -- of the stars targeted
for analysis.

\begin{figure}
\centering
\includegraphics{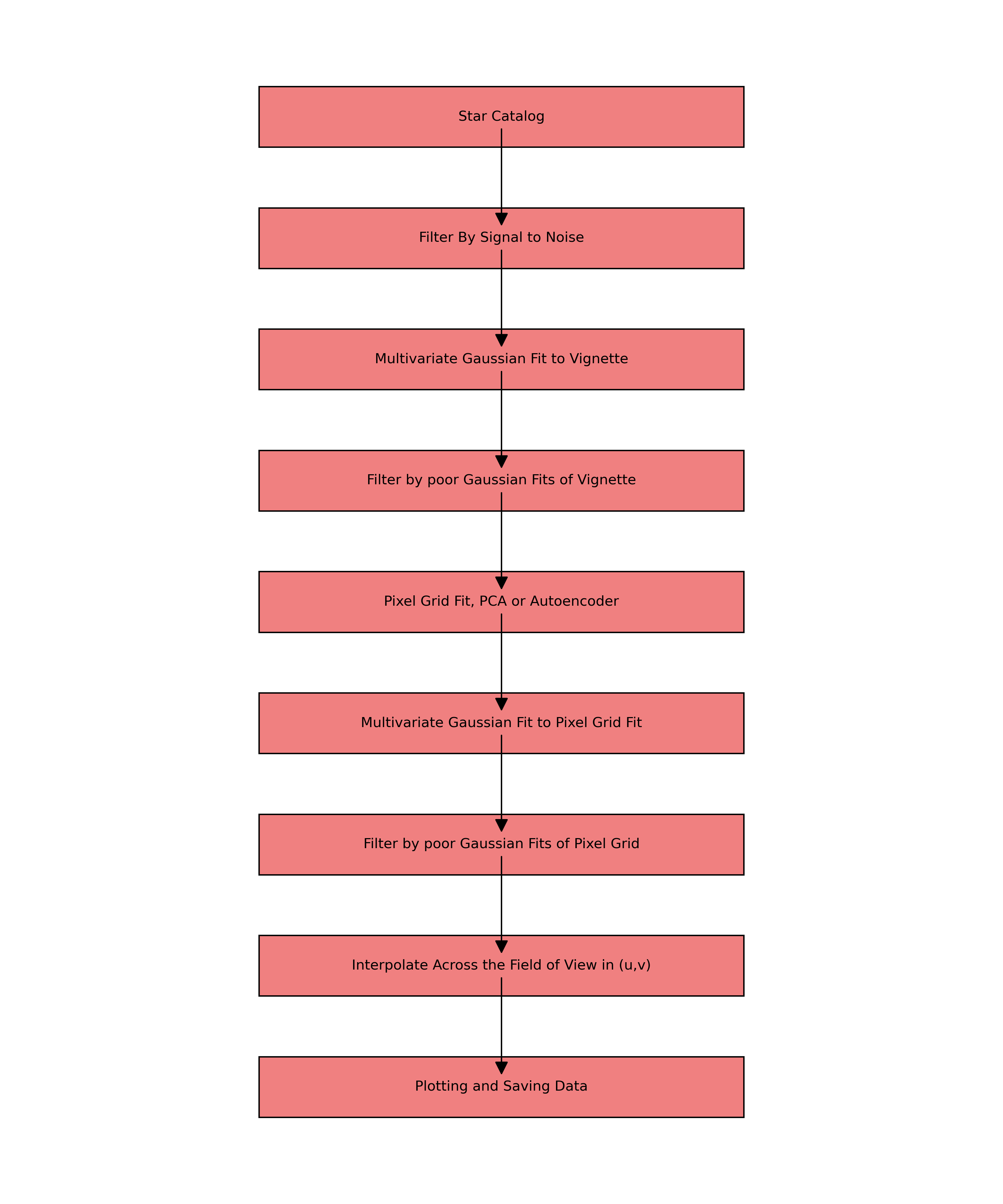}
\caption{ShOpt Workflow}
\end{figure}

\hypertarget{statement-of-need}{%
\section{Statement of need}\label{statement-of-need}}

Empirical PSF characterization tools like Point Spread Function
Extractor (PSFEx (\protect\hyperlink{ref-2011ASPC}{Bertin, 2011})) and
Point Spread Functions in the Full Field of View (PIFF
(\protect\hyperlink{ref-Jarvis_2020}{Jarvis et al., 2020})) are widely
popular in astrophysics. However, the quality of PIFF and PSFEx models
tends to be quite sensitive to the parameter values used to run the
software, with optimization sometimes relying on brute-force
guess-and-check runs. PIFF is also notably inefficient for large,
well-sampled images, taking hours in the worst cases. The JWST's Near
Infrared Camera (NIRCam) offers vast scientific opportunities (e.g.
(\protect\hyperlink{ref-casey2023cosmosweb}{Casey, Kartaltepe, et al.,
2023})); at the same time, this unprecedented data brings new challenges
for PSF modeling:

\begin{enumerate}
\def\labelenumi{(\arabic{enumi})}
\item
  Analytic functions like Gaussians are incomplete descriptions of the
  NIRCam PSF, as evident from Figures 2 and 3. This calls for
  well-thought-out, non-parametric modeling and diagnostic tools that
  can capture the full dynamic range of the NIRCam PSF. \texttt{ShOpt}
  provides these models and diagnostics out of the box.
\item
  The NIRCam detectors have pixel scales of 0.03 (short wavelength
  channel) and 0.06 (long wavelength channel) arcseconds per pixel
  (\protect\hyperlink{ref-BSPIE}{Beichman et al., 2012};
  \protect\hyperlink{ref-10.1117ux2f12.489103}{Rieke et al., 2003},
  \protect\hyperlink{ref-20052005SPIE}{2005}). At these pixel scales,
  star vignettes need to be at least \(131\) by \(131\) pixels across to
  fully capture the wings of the PSFs (4-5 arcseconds). These vignette
  sizes are 3-5 times larger than the ones used in surveys such as DES
  (\protect\hyperlink{ref-Jarvis_2020}{Jarvis et al., 2020}) and
  SuperBIT (\protect\hyperlink{ref-mccleary2023lensing}{McCleary et al.,
  2023}) and force us to evaluate how well existing PSF fitters scale to
  this size. \texttt{ShOpt} has been designed for computational
  efficiency and aims to meet the requirements of detectors like NIRCam.
\end{enumerate}

\begin{figure}
\centering
\includegraphics{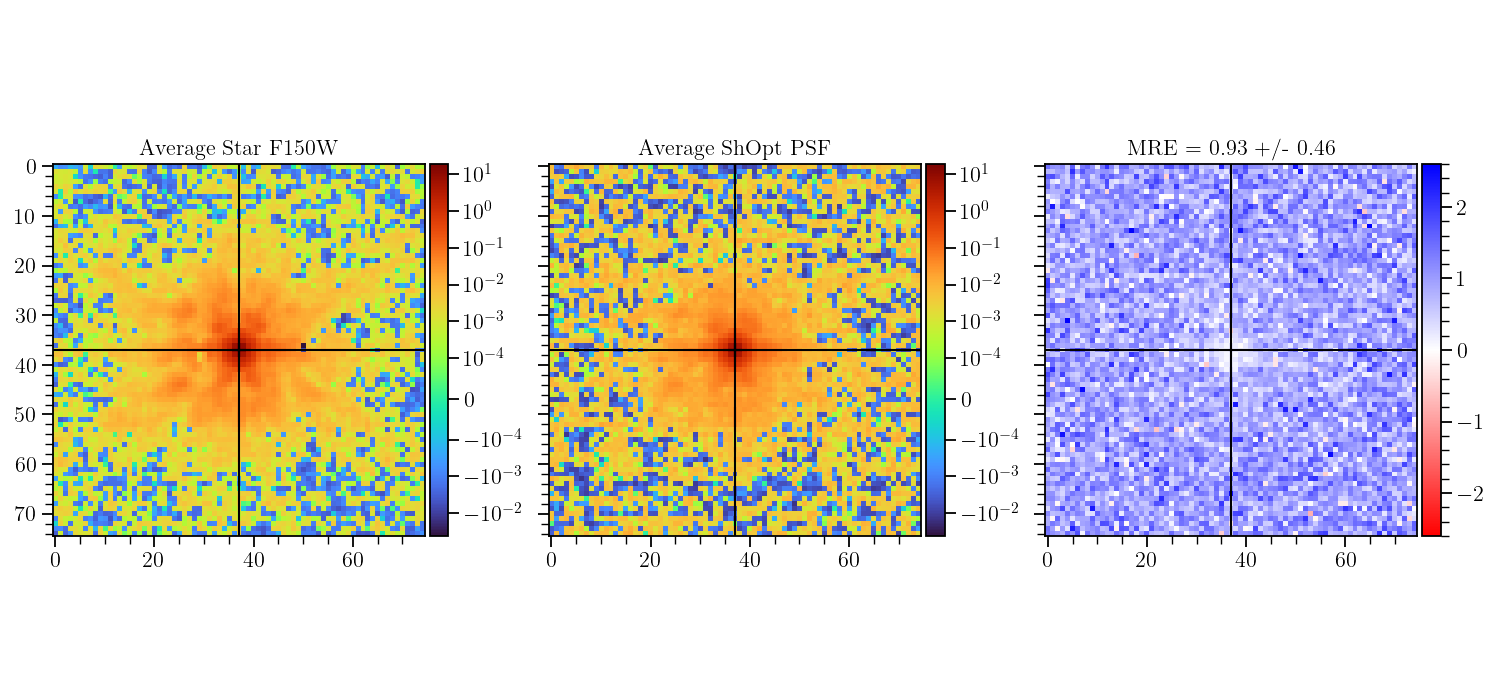}
\caption{The plot on the left shows the average cutout of all stars in a
supplied catalog. The plot in the middle shows the average point spread
function model for each star. The plot on the right shows the average
normalized error between the observed star cutouts and the point spread
function model.}
\end{figure}

\begin{figure}
\centering
\includegraphics{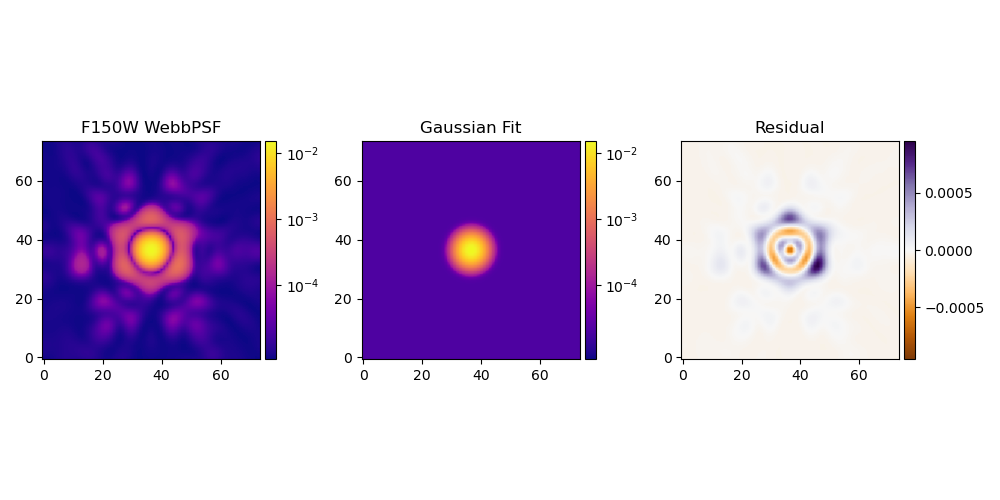}
\caption{The plot on the left shows an idealized forward model of the
NIRCam F150W PSF made using WebbPSF. The middle shows a Gaussian
approximation to the PSF. The right shows the residual between the
WebbPSF model and the Gaussian approximation.}
\end{figure}

\texttt{ShOpt} bridges the speed of \texttt{PSFex} with the features of
\texttt{PIFF} using fewer configurable hyperparameters. \texttt{ShOpt}
employs a myriad of techniques to optimize the speed of the program.
First and foremost, \texttt{ShOpt} is equipped with support for
multithreading. Polynomial interpolation is used for handling PSF
variations across the field of view. The polynomials given to each basis
element of the PSF are independent of one another and therefore can be
distributed to different CPU threads to be run in parallel.
\texttt{ShOpt} also introduces new methods for fitting both analytic
profiles and pixel based profiles. If an analytic profile is used to
model the PSF, then there are \(3\) basis elements parameterizing the
model. We choose 2D elliptical Gaussians for these analytic profiles
because they are cheap to compute. Moreover, we use the Limited Memory
Broyden--Fletcher--Goldfarb--Shanno algorithm (LBFGS) to find these
paramters. This is faster but more memory intensive than Conjugate
Gradient, the algorithm used in \texttt{PIFF}. Moreover, the \(3\) basis
elements are constrained to the manifold \(B_2(r) \times \mathbb{R}_+\).
We constructed a function that maps any point in \(\mathbb{R}^3\) into
\(B_2(r) \times \mathbb{R}_+\). The LBFGS algorithm uses successive
iterations to converge to a solution for the \(3\) basis elements, and
so we use this function to ensure that our update steps in LBFGS do not
leave the constraint. For the pixel basis, both \texttt{PIFF} and
\texttt{PSFex} approximate the PSF by minimizing the reduced \(\chi^2\)
between a grid of pixels and a star vignette. PCA can quickly achieve
the same purpose of approximating the input vignette without overfitting
to background noise. We also provide an autoencoder mode, which uses
deep learning to reconstruct the image. The weights and biases are not
reset between stars, so the knowledge of how to reconstruct one star is
transfered to the next. This in turn leads to fewer training iterations.
Finally, \texttt{ShOpt} is written in Julia. Julia uses a just in time
compiler which makes it faster than intepreted languages such as Python
and has shown to be a good choice for performance critical code
(\protect\hyperlink{ref-Stanitzki_2021}{Stanitzki \& Strube, 2021}).

\hypertarget{state-of-the-field}{%
\section{State of the Field}\label{state-of-the-field}}

The JWST captures images at high resolution and at wavelengths of light
that have been previously unexplored
(\protect\hyperlink{ref-Gardner_2006}{Gardner et al., 2006}). With these
images we are seeing farther into the early universe than we ever have
before. The difficulties of producing good PSF models for the JWST are
emblematic of a larger problem: Our data sets are getting bigger and
existing software was not built to scale. That is to say, the
advancements in software are falling behind the advances in
instrumentation. Not only does \texttt{ShOpt} produces PSF models for
JWST NIRCam images, it also sets the precedent for designing software
that scales.

There are several existing empirical PSF fitters, in addition to a
forward model of the JWST PSFs developed by STScI
(\protect\hyperlink{ref-Jarvis_2020}{Jarvis et al., 2020} ;
\protect\hyperlink{ref-2011ASPC}{Bertin, 2011};
\protect\hyperlink{ref-2014SPIE}{Perrin et al., 2014} ;
\protect\hyperlink{ref-2012SPIE}{Perrin et al., 2012}). We describe them
here and draw attention to their strengths and weaknesses to further
motivate the development of \texttt{ShOpt.jl}. As described in the
statement of need, \texttt{PSFex} was one of the first precise and
general purpose tools used for empirical PSF fitting. However, the Dark
Energy Survey collaboration reported small but noticeable discrepancies
between the sizes of \texttt{PSFex} models and the sizes of observed
stars. They also reported ripple-like patterns in the spatial variation
of star-PSF residuals across the field of view
(\protect\hyperlink{ref-Jarvis_2020}{Jarvis et al., 2020}), which they
attributed to the astrometric distortion solutions for the Dark Energy
Camera.

These findings motivated the Dark Energy Survey's development of
\texttt{PIFF} (Point Spread Functions in the Full Field of View).
\texttt{PIFF} works in sky coordinates on the focal plane, as opposed to
image pixel coordinates used in \texttt{PSFex}, which minimized the
ripple patterns in the star-PSF residuals and the PSF model size bias.
(Based on the DES findings, \texttt{ShOpt} also works directly in sky
coordinates.) \texttt{PIFF} is written in Python, a language with a
large infrastructure for astronomical data analysis, for example Astropy
(\protect\hyperlink{ref-2022ApJ}{Astropy Collaboration et al., 2022})
and Galsim (\protect\hyperlink{ref-rowe2015galsim}{Rowe et al., 2015}).
The choice of language makes \texttt{PIFF} software more accessible to
programmers in the astrophysics community than \texttt{PSFex}, which was
first written in \texttt{C} in 2007 and much less approachable for a
community of open source developers. One of the motivations of
\texttt{ShOpt} was to write astrophysics specific software in
\texttt{Julia}, because \texttt{Julia} provides a good balance of
readability and speed with its high-level functional paradigm and
just-in-time compiler (\protect\hyperlink{ref-Stanitzki_2021}{Stanitzki
\& Strube, 2021}). Julia ranks behind Python, IDL, Matlab, and Fortran
in full-text mentions in astronomical literature
(\protect\hyperlink{ref-TheAstropyCollaboration_2022}{Collaboration et
al., 2022}). We are optimistic that \texttt{ShOpt} will demonstrate that
Julia is an appealing choice for programming in astronomy despite its
low adoption to date. There is also recent work on using \texttt{PSFr}
for PSF reconstructions that has been applied to JWST data
(\protect\hyperlink{ref-2022ApJ939L28D}{Ding et al., 2022};
\protect\hyperlink{ref-2022ApJ938L14M}{Merlin et al., 2022};
\protect\hyperlink{ref-2023ApJ942L27S}{Santini et al., 2023};
\protect\hyperlink{ref-2022ApJ938L17Y}{Yang et al., 2022}). Similar to
\texttt{ShOpt}, \texttt{PSFr} begins its PSF modeling by stacking stars
to form an initial guess. However, rather than using polynomial
interpolation to address spatial variations, \texttt{PSFr} employs an
iterative process of shifting the pixels in its PSF model. This process
continues until the model can adequately represent all of the stars in
the catalog. Finally, the PSF fitter \texttt{STARRED}
(\protect\hyperlink{ref-michalewicz2023starred}{Michalewicz et al.,
2023}) has been shown to produce PSF models competitive with
\texttt{PSFr} and \texttt{PSFex}. Like \texttt{ShOpt}, \texttt{STARRED}
puts an emphasis on computational efficiency and uses the JAX package
(\protect\hyperlink{ref-schoenholz2019jax}{Schoenholz \& Cubuk, 2019})
to achieve just-in-time compilation. There is future work to be done to
benchmarking all of these empirical approaches on JWST NIRCam data.

While WebbPSF provides highly precise forward models of the JWST PSF,
these models are defined for single-epoch exposures
(\protect\hyperlink{ref-2014SPIE}{Perrin et al., 2014},
\protect\hyperlink{ref-2012SPIE}{2012}). Much of the NIRCam science is
accomplished with image mosaics -- essentially, the combination of
single exposure detector images into a larger, deeper image. The
rotation of the camera between exposures, the astrometric
transformations and resampling of images before their combination into a
mosaic, and the mosaic's large area all make the application of WebbPSF
models to mosaics a non-trivial procedure. This has been done quite
effectively (\protect\hyperlink{ref-ji2023jades}{Ji et al., 2023}),
however, it is not as easy to reproduce compared to running an empirical
characterization tool like \texttt{ShOpt}.

As outlined in the state of the field, \texttt{ShOpt} is a tool built
with the user experience in mind that attempts to bridge the strengths
of existing PSF fitters. \texttt{ShOpt}'s combination of speed, user
friendliness, and accuracy enable the science goals of the COSMOS-Web
survey, detailed below.

The COMOS-Web survey is the largest cycle 1 JWST extragalactic survey
according to area and prime time allocation
(\protect\hyperlink{ref-casey2023cosmosweb}{Casey, Kartaltepe, et al.,
2023}), and covers \(0.54 ~deg^2\)
(\protect\hyperlink{ref-BSPIE}{Beichman et al., 2012};
\protect\hyperlink{ref-Rieke_2023}{Rieke et al., 2023}). Among other
science goals, the COMOS-Web survey will use the JWST to detect
thousands of galaxies in the Epoch of Reionization \((6 \sim z \sim 11)\) to
create one of the highest resolution large scale structure maps of the
early universe (\protect\hyperlink{ref-casey2023cosmosweb}{Casey,
Kartaltepe, et al., 2023}). JWST data has also been used to pick out
active galactic nuclei from host galaxies
(\protect\hyperlink{ref-zhuang2023characterization}{Zhuang \& Shen,
2023}) and indentify \(15\) candidate galaxies whose luminosities push
the limits of our \(\Lambda\)CDM galaxy formation models
(\protect\hyperlink{ref-franco2023cosmosweb}{Casey, Akins, et al.,
2023}). These science cases all rely upon good PSF modeling and
underscore the importance of \texttt{ShOpt}.

\hypertarget{acknowledgements}{%
\section{Acknowledgements}\label{acknowledgements}}

This material is based upon work supported by a Northeastern University
Undergraduate Research and Fellowships PEAK Experiences Award. E.B. was
also supported by a Northeastern University Physics Department Co-op
Research Fellowship. Support for COSMOS-Web was provided by NASA through
grant JWST-GO-01727 and HST-AR-15802 awarded by the Space Telescope
Science Institute, which is operated by the Association of Universities
for Research in Astronomy, Inc., under NASA contract NAS 5-26555. This
work was made possible by utilizing the CANDIDE cluster at the Institut
d'Astrophysique de Paris. Further support was provided by Research
Computers at Northeastern University. Additionally, E.B. thanks
Professor David Rosen for giving some valuable insights during the early
stages of this work. The authors gratefully acknowledge the use of
simulated and real data from the COSMOS-Web survey in developing ShOpt,
as well as many conversations with COSMOS-Web scientists.

\hypertarget{references}{%
\section*{References}\label{references}}
\addcontentsline{toc}{section}{References}

\hypertarget{refs}{}
\begin{CSLReferences}{1}{0}
\leavevmode\vadjust pre{\hypertarget{ref-2022ApJ}{}}%
Astropy Collaboration, Price-Whelan, A. M., Lim, P. L., Earl, N.,
Starkman, N., Bradley, L., Shupe, D. L., Patil, A. A., Corrales, L.,
Brasseur, C. E., Nöthe, M., Donath, A., Tollerud, E., Morris, B. M.,
Ginsburg, A., Vaher, E., Weaver, B. A., Tocknell, J., Jamieson, W.,
\ldots{} Astropy Project Contributors. (2022). {The Astropy Project:
Sustaining and Growing a Community-oriented Open-source Project and the
Latest Major Release (v5.0) of the Core Package}. \emph{935}(2), 167.
\url{https://doi.org/10.3847/1538-4357/ac7c74}

\leavevmode\vadjust pre{\hypertarget{ref-BSPIE}{}}%
Beichman, C. A., Rieke, M., Eisenstein, D., Greene, T. P., Krist, J.,
McCarthy, D., Meyer, M., \& Stansberry, J. (2012). {Science
opportunities with the near-IR camera (NIRCam) on the James Webb Space
Telescope (JWST)}. In M. C. Clampin, G. G. Fazio, H. A. MacEwen, \& Jr.
Oschmann Jacobus M. (Eds.), \emph{Space telescopes and instrumentation
2012: Optical, infrared, and millimeter wave} (Vol. 8442, p. 84422N).
\url{https://doi.org/10.1117/12.925447}

\leavevmode\vadjust pre{\hypertarget{ref-2011ASPC}{}}%
Bertin, E. (2011). {Automated Morphometry with SExtractor and PSFEx}. In
I. N. Evans, A. Accomazzi, D. J. Mink, \& A. H. Rots (Eds.),
\emph{Astronomical data analysis software and systems XX} (Vol. 442, p.
435).

\leavevmode\vadjust pre{\hypertarget{ref-franco2023cosmosweb}{}}%
Casey, C. M., Akins, H. B., Shuntov, M., Ilbert, O., Paquereau, L.,
Franco, M., Hayward, C. C., Finkelstein, S. L., Boylan-Kolchin, M.,
Robertson, B. E., Allen, N., Brinch, M., Cooper, O. R., Ding, X.,
Drakos, N. E., Faisst, A. L., Fujimoto, S., Gillman, S., Harish, S.,
\ldots{} Zavala, J. A. (2023). \emph{COSMOS-web: Intrinsically luminous
z\(\gtrsim\)10 galaxy candidates test early stellar mass assembly}.
\url{https://arxiv.org/abs/2308.10932}

\leavevmode\vadjust pre{\hypertarget{ref-casey2023cosmosweb}{}}%
Casey, C. M., Kartaltepe, J. S., Drakos, N. E., Franco, M., Harish, S.,
Paquereau, L., Ilbert, O., Rose, C., Cox, I. G., Nightingale, J. W.,
Robertson, B. E., Silverman, J. D., Koekemoer, A. M., Massey, R.,
McCracken, H. J., Rhodes, J., Akins, H. B., Amvrosiadis, A.,
Arango-Toro, R. C., \ldots{} Zavala, J. A. (2023). \emph{COSMOS-web: An
overview of the JWST cosmic origins survey}.
\url{https://arxiv.org/abs/2211.07865}

\leavevmode\vadjust pre{\hypertarget{ref-TheAstropyCollaboration_2022}{}}%
Collaboration, T. A., Price-Whelan, A. M., Lim, P. L., Earl, N.,
Starkman, N., Bradley, L., Shupe, D. L., Patil, A. A., Corrales, L.,
Brasseur, C. E., Nöthe, M., Donath, A., Tollerud, E., Morris, B. M.,
Ginsburg, A., Vaher, E., Weaver, B. A., Tocknell, J., Jamieson, W.,
\ldots{} Contributors, A. P. (2022). The astropy project: Sustaining and
growing a community-oriented open-source project and the latest major
release (v5.0) of the core package*. \emph{The Astrophysical Journal},
\emph{935}(2), 167. \url{https://doi.org/10.3847/1538-4357/ac7c74}

\leavevmode\vadjust pre{\hypertarget{ref-2022ApJ939L28D}{}}%
Ding, X., Silverman, J. D., \& Onoue, M. (2022). {Opening the Era of
Quasar-host Studies at High Redshift with JWST}. \emph{939}(2), L28.
\url{https://doi.org/10.3847/2041-8213/ac9c02}

\leavevmode\vadjust pre{\hypertarget{ref-Gardner_2006}{}}%
Gardner, J. P., Mather, J. C., Clampin, M., Doyon, R., Greenhouse, M.
A., Hammel, H. B., Hutchings, J. B., Jakobsen, P., Lilly, S. J., Long,
K. S., Lunine, J. I., Mccaughrean, M. J., Mountain, M., Nella, J.,
Rieke, G. H., Rieke, M. J., Rix, H.-W., Smith, E. P., Sonneborn, G.,
\ldots{} Wright, G. S. (2006). The james webb space telescope.
\emph{Space Science Reviews}, \emph{123}(4), 485--606.
\url{https://doi.org/10.1007/s11214-006-8315-7}

\leavevmode\vadjust pre{\hypertarget{ref-Jarvis_2020}{}}%
Jarvis, M., Bernstein, G. M., Amon, A., Davis, C., Lé get, P. F.,
Bechtol, K., Harrison, I., Gatti, M., Roodman, A., Chang, C., Chen, R.,
Choi, A., Desai, S., Drlica-Wagner, A., Gruen, D., Gruendl, R. A.,
Hernandez, A., MacCrann, N., Meyers, J., \ldots{} and, R. D. W. (2020).
Dark energy survey year 3 results: Point spread function modelling.
\emph{Monthly Notices of the Royal Astronomical Society}, \emph{501}(1),
1282--1299. \url{https://doi.org/10.1093/mnras/staa3679}

\leavevmode\vadjust pre{\hypertarget{ref-ji2023jades}{}}%
Ji, Z., Williams, C. C., Tacchella, S., Suess, K. A., Baker, W. M.,
Alberts, S., Bunker, A. J., Johnson, B. D., Robertson, B., Sun, F.,
Eisenstein, D. J., Rieke, M., Maseda, M. V., Hainline, K., Hausen, R.,
Rieke, G., Willmer, C. N. A., Egami, E., Shivaei, I., \ldots{} Sandles,
L. (2023). \emph{JADES + JEMS: A detailed look at the buildup of central
stellar cores and suppression of star formation in galaxies at redshifts
3 \textless{} z \textless{} 4.5}. \url{https://arxiv.org/abs/2305.18518}

\leavevmode\vadjust pre{\hypertarget{ref-mccleary2023lensing}{}}%
McCleary, J. E., Everett, S. W., Shaaban, M. M., Gill, A. S.,
Vassilakis, G. N., Huff, E. M., Massey, R. J., Benton, S. J., Brown, A.
M., Clark, P., \& others. (2023). Lensing in the blue II: Estimating the
sensitivity of stratospheric balloons to weak gravitational lensing.
\emph{arXiv Preprint arXiv:2307.03295}.

\leavevmode\vadjust pre{\hypertarget{ref-2022ApJ938L14M}{}}%
Merlin, E., Bonchi, A., Paris, D., Belfiori, D., Fontana, A.,
Castellano, M., Nonino, M., Polenta, G., Santini, P., Yang, L.,
Glazebrook, K., Treu, T., Roberts-Borsani, G., Trenti, M., Birrer, S.,
Brammer, G., Grillo, C., Calabrò, A., Marchesini, D., \ldots{} Wang, X.
(2022). {Early Results from GLASS-JWST. II. NIRCam Extragalactic Imaging
and Photometric Catalog}. \emph{938}(2), L14.
\url{https://doi.org/10.3847/2041-8213/ac8f93}

\leavevmode\vadjust pre{\hypertarget{ref-michalewicz2023starred}{}}%
Michalewicz, K., Millon, M., Dux, F., \& Courbin, F. (2023). STARRED: A
two-channel deconvolution method with starlet regularization.
\emph{Journal of Open Source Software}, \emph{8}(85), 5340.

\leavevmode\vadjust pre{\hypertarget{ref-2014SPIE}{}}%
Perrin, M. D., Sivaramakrishnan, A., Lajoie, C.-P., Elliott, E., Pueyo,
L., Ravindranath, S., \& Albert, Loic. (2014). {Updated point spread
function simulations for JWST with WebbPSF}. In Jr. Oschmann Jacobus M.,
M. Clampin, G. G. Fazio, \& H. A. MacEwen (Eds.), \emph{Space telescopes
and instrumentation 2014: Optical, infrared, and millimeter wave} (Vol.
9143, p. 91433X). \url{https://doi.org/10.1117/12.2056689}

\leavevmode\vadjust pre{\hypertarget{ref-2012SPIE}{}}%
Perrin, M. D., Soummer, R., Elliott, E. M., Lallo, M. D., \&
Sivaramakrishnan, A. (2012). {Simulating point spread functions for the
James Webb Space Telescope with WebbPSF}. In M. C. Clampin, G. G. Fazio,
H. A. MacEwen, \& Jr. Oschmann Jacobus M. (Eds.), \emph{Space telescopes
and instrumentation 2012: Optical, infrared, and millimeter wave} (Vol.
8442, p. 84423D). \url{https://doi.org/10.1117/12.925230}

\leavevmode\vadjust pre{\hypertarget{ref-10.1117ux2f12.489103}{}}%
Rieke, M. J., Baum, S. A., Beichman, C. A., Crampton, D., Doyon, R.,
Eisenstein, D., Greene, T. P., Hodapp, K.-W., Horner, S. D., Johnstone,
D., Lesyna, L., Lilly, S., Meyer, M., Martin, P., Jr., D. W. M., Rieke,
G. H., Roellig, T. L., Stauffer, J., Trauger, J. T., \& Young, E. T.
(2003). {NGST NIRCam scientific program and design concept}. In J. C.
Mather (Ed.), \emph{IR space telescopes and instruments} (Vol. 4850, pp.
478--485). International Society for Optics; Photonics; SPIE.
\url{https://doi.org/10.1117/12.489103}

\leavevmode\vadjust pre{\hypertarget{ref-Rieke_2023}{}}%
Rieke, M. J., Kelly, D. M., Misselt, K., Stansberry, J., Boyer, M.,
Beatty, T., Egami, E., Florian, M., Greene, T. P., Hainline, K.,
Leisenring, J., Roellig, T., Schlawin, E., Sun, F., Tinnin, L.,
Williams, C. C., Willmer, C. N. A., Wilson, D., Clark, C. R., \ldots{}
Young, E. T. (2023). Performance of NIRCam on JWST in flight.
\emph{Publications of the Astronomical Society of the Pacific},
\emph{135}(1044), 028001. \url{https://doi.org/10.1088/1538-3873/acac53}

\leavevmode\vadjust pre{\hypertarget{ref-20052005SPIE}{}}%
Rieke, M. J., Kelly, D., \& Horner, S. (2005). {Overview of James Webb
Space Telescope and NIRCam's Role}. In J. B. Heaney \& L. G. Burriesci
(Eds.), \emph{Cryogenic optical systems and instruments XI} (Vol. 5904,
pp. 1--8). \url{https://doi.org/10.1117/12.615554}

\leavevmode\vadjust pre{\hypertarget{ref-rowe2015galsim}{}}%
Rowe, B., Jarvis, M., Mandelbaum, R., Bernstein, G. M., Bosch, J.,
Simet, M., Meyers, J. E., Kacprzak, T., Nakajima, R., Zuntz, J.,
Miyatake, H., Dietrich, J. P., Armstrong, R., Melchior, P., \& Gill, M.
S. S. (2015). \emph{GalSim: The modular galaxy image simulation
toolkit}. \url{https://arxiv.org/abs/1407.7676}

\leavevmode\vadjust pre{\hypertarget{ref-2023ApJ942L27S}{}}%
Santini, P., Fontana, A., Castellano, M., Leethochawalit, N., Trenti,
M., Treu, T., Belfiori, D., Birrer, S., Bonchi, A., Merlin, E., Mason,
C., Morishita, T., Nonino, M., Paris, D., Polenta, G., Rosati, P., Yang,
L., Boyett, K., Bradac, M., \ldots{} Wang, X. (2023). {Early Results
from GLASS-JWST. XI. Stellar Masses and Mass-to-light Ratio of z
\textgreater{} 7 Galaxies}. \emph{942}(2), L27.
\url{https://doi.org/10.3847/2041-8213/ac9586}

\leavevmode\vadjust pre{\hypertarget{ref-schoenholz2019jax}{}}%
Schoenholz, S. S., \& Cubuk, E. D. (2019). \emph{Jax md: End-to-end
differentiable, hardware accelerated, molecular dynamics in pure
python}.

\leavevmode\vadjust pre{\hypertarget{ref-Stanitzki_2021}{}}%
Stanitzki, M., \& Strube, J. (2021). Performance of julia for high
energy physics analyses. \emph{Computing and Software for Big Science},
\emph{5}(1). \url{https://doi.org/10.1007/s41781-021-00053-3}

\leavevmode\vadjust pre{\hypertarget{ref-2022ApJ938L17Y}{}}%
Yang, L., Morishita, T., Leethochawalit, N., Castellano, M., Calabrò,
A., Treu, T., Bonchi, A., Fontana, A., Mason, C., Merlin, E., Paris, D.,
Trenti, M., Roberts-Borsani, G., Bradac, M., Vanzella, E., Vulcani, B.,
Marchesini, D., Ding, X., Nanayakkara, T., \ldots{} Wang, X. (2022).
{Early Results from GLASS-JWST. V: The First Rest-frame Optical
Size-Luminosity Relation of Galaxies at z \textgreater{} 7}.
\emph{938}(2), L17. \url{https://doi.org/10.3847/2041-8213/ac8803}

\leavevmode\vadjust pre{\hypertarget{ref-zhuang2023characterization}{}}%
Zhuang, M.-Y., \& Shen, Y. (2023). \emph{Characterization of JWST NIRCam
PSFs and implications for AGN+host image decomposition}.
\url{https://arxiv.org/abs/2304.13776}

\end{CSLReferences}

\end{document}